\documentclass[conference]{IEEEtran}
\usepackage{blindtext, graphicx}
\usepackage{lineno}
\usepackage{amsmath}
\usepackage{amssymb}
\usepackage{mathtools}
\usepackage{textcomp}
\usepackage{caption}
\usepackage{subcaption}

\usepackage{algcompatible}
\usepackage{algorithm}

\usepackage{array}
\usepackage{multirow}
\usepackage{booktabs}

\usepackage[table,xcdraw]{xcolor}
\usepackage{tablefootnote}
\usepackage{tabularx}

\usepackage{makecell}
\usepackage[pdftex,colorlinks,citecolor=blue]{hyperref}

\begin{document}

\title{ Scheduling Bag-of-Tasks in Clouds using  Spot  and  Burstable Virtual Machines}

\author{\IEEEauthorblockN{Luan Teylo}
\IEEEauthorblockA{Institute of Computing\\
Fluminense Federal University\\
Niter{\'o}i, Brasil 24210-346\\
Email: luanteylo@ic.uff.br \\
}

\and
\IEEEauthorblockN{Luciana Arantes}
\IEEEauthorblockA{LIP6\\
Sorbonne Univ., CNRS, Inria\\
Paris, France\\
Email: luciana.arantes@lip6.fr}

\and
\IEEEauthorblockN{Pierre Sens}
\IEEEauthorblockA{LIP6\\
Sorbonne Univ., CNRS, Inria\\
Paris, France\\
Email: pierre.sens@lip6.fr}

\and
\IEEEauthorblockN{L{\'u}cia Maria de A. Drummond}
\IEEEauthorblockA{Institute of Computing\\
Fluminense Federal University\\
Niter{\'o}i, Brasil 24210-346\\
Email: lucia@ic.uff.br}
}

\maketitle
\thispagestyle{plain}
\pagestyle{plain}

\begin{abstract}

  Leading Cloud providers offer several types of Virtual Machines (VMs) in diverse contract models, with different guarantees in terms of availability and reliability. Among them, the most popular contract models are the on-demand and the spot models. In the former, on-demand VMs are allocated for a fixed cost per time unit, and their availability is ensured during the whole execution. On the other hand, in the spot market, VMs are offered with a huge discount when compared to the on-demand VMs, but their availability fluctuates according to the cloud's current demand that can terminate or hibernate a spot VM at any time. Furthermore, in order to cope with workload variations, cloud providers have also introduced the concept of burstable VMs which are able to burst up their respective baseline CPU performance during a limited period of time with an up to 20\% discount when compared to an equivalent non-burstable on-demand VMs. In the current work, we present the Burst Hibernation-Aware Dynamic Scheduler (Burst-HADS), a framework that schedules and executes tasks of  Bag-of-Tasks applications with deadline constraints by exploiting spot and on-demand burstable VMs, aiming at minimizing both the monetary cost and the execution time. Based on  ILS metaheuristics, Burst-HADS defines an initial scheduling map of tasks to VMs which can then be dynamically altered by migrating tasks of a hibernated spot VM or by performing work-stealing when VMs become idle. Performance results on Amazon EC2 cloud with different applications show that, when compared to a solution that uses only regular on-demand instances, Burst-HADS reduces the monetary cost of the execution and meet the application deadline even in scenarios with high spot hibernation rates. It also reduces the total execution time when compared to a solution that uses only spot and non-burstable on-demand instances. 

\end{abstract}


\IEEEpeerreviewmaketitle

\section{Introduction}
\label{sec:introduction}

In the past few years, cloud computing has emerged as an attractive option to execute different applications due to several advantages that it brings compared with dedicated infrastructure. Clouds provide a significant reduction in operational costs, besides offering rapid elastic provisioning of computing resources like virtual machines (VMs) and storage. However, in cloud environments,  besides the usual goal of minimizing the application's execution time,  it is also essential to reduce the monetary cost of using its resources, i.e., there is a trade-off between performance and budget.

Most cloud platforms enable users to dynamically acquire computational resources wrapped as virtual machines (VMs).  The users can reserve the VMs depending on their application requirements (CPU, memory, I/O, etc.) in a pay-as-you-use price model. Furthermore, cloud providers offer VMs in different contract models, with different guarantees in terms of availability and volatility. For instance, in Amazon EC2, there are three main contract models (also call markets): i) {\it reserved} market, where the user pays an upfront price, guaranteeing long-term availability; ii) {\it on-demand} market which is allocated for specific periods of time, and incurs a fixed cost per unit time of use, ensuring the availability of the instance during this period; iii) {\it spot} market in which unused resources are available up to 90\% discount when compared to the on-demand model.

In the three markets, there exist a wide range of VM types that suit different user requirements. According to Amazon Web Service (AWS), ``Instance types comprise varying combinations of virtual CPUs (vCPUs), memory, storage, and networking capacity and give you the flexibility to choose the appropriate mix of resources for your applications. Each instance type includes one or more instance sizes, allowing you to scale your resources to the requirements of your target workload”\footnote{\url{https://aws.amazon.com/ec2/instance-types}}. Instances types are grouped into families based on their use case. For example, the compute-optimized instances (C3, C4, and C5) are ideal for compute-bound applications that require high-performance processors. 

Regarding the spot market, the availability of its VMs fluctuates according to the cloud's current demand. If there are not enough resources to meet clients' requests, the cloud provider can interrupt a spot VM (temporarily or definitively). Despite the risk of unavailability, the main advantage of spot VMs is that their costs are much lower than on-demand VMs since the user requests unused instances at steep discounts. An interrupted spot VM instance can either terminate or hibernate.  If the VM will be terminated, the cloud provider warns the user two minutes before its interruption. On the other hand, hibernated VM instances are frozen immediately after noticing the user. In this case, EC2 saves the VM instance memory and context in the root of EC2 Block Storage (EBS) volume, and during the VM's interruption period, the user is only charged for the EBS storage use. EC2 resumes the hibernated spot instance, reloading the saved memory and context, only when there is enough available resource whose price is lower than the maximum one, which the user agreed to be charged.

Beside the markets, all leading cloud providers introduced in the last years the concept of burstable VM that can sprint its performance during a limited period of time to cope with sudden workload variations. By operating on a CPU credit regime that controls the processing power offered to users, burstable VM instances can use 100\% of the VM's processing power (burst mode) or only a fraction of that processing power  (baseline mode), depending on such credits.  They accumulate CPU credits per hour, whose amount depends on the instance type. If a burstable instance uses fewer CPU resources than required for baseline performance (for example,  when it is idle), the unspent CPU credits are accrued in the CPU credit balance of the instance. If a burstable VM needs to burst above the baseline performance level, it spends the accrued credits. The more credits that a burstable performance instance has accrued, the longer it can burst beyond its baseline when higher performance is required.
Burstable instances have two main advantages: i) they are offered with an up to 20\% discount compared to non-burstable on-demand instances with equivalent computational resources, and ii) contrarily to spot VMs, they are not prone to revocation. On the other hand, to obtain monetary advantages of burstable instances, the user has to control their respective CPU credit usage by monitoring their baseline performance and bursting during their entire running.  In Amazon EC2, a burstable VMs are of type T3, T3a, or T2.

This paper presents a scheduler for Bag-of-Tasks applications with deadline constraints, exploring hibernate-prone spot instances and on-demand burstable instances aiming at minimizing both the monetary cost and the execution time of applications.
The proposed scheduler, denoted Burst Hibernation-Aware Dynamic Scheduler (Burst-HADS), is an event-driven scheduling framework built in a modular way with two main scheduling modules: i) the {\bf Primary Scheduling Module} that defines an initial scheduling map of tasks to VMs; and ii) a {\bf Dynamic Scheduling Module} responsible for task migration or task rollback recovering to/in idle VMs.  It also provides a {\bf Fault Tolerance Module} that takes checkpoints periodically, enabling that a task, when migrated, re-starts from the last checkpoint.

In a previous work \cite{teylocc}, we had already explored the use of hibernation-prone spot instances to minimize the monetary cost of BoT applications execution, respecting their respective deadline constraints. To meet such a deadline, even in the presence of multiple hibernations,  new on-demand VMs, not allocated in the initial mapping, might be dynamically launched. In this case, tasks of the hibernated spot instances and those not executed yet are migrated to on-demand instances.  Although the strategy can significantly reduce the monetary costs, always respecting the application deadline, if VM spots hibernate, the application's total execution time might considerably increase when compared with a solution that uses only on-demand instances. To tackle such a problem, in this article, we are interested in investigating how burstable instances can be used to reduce the impact in the execution time of the application and the corresponding monetary costs. Several scenarios are studied and evaluated, showing that the proposed burstable instance approach can be very advantageous, guaranteeing the application's deadline without degrading application execution time, even in the presence of hibernations, and still optimizing the monetary costs.

The main contributions of the paper are:

\begin{itemize}
     \item A mathematical formulation of the static scheduling  problem;
     \item An Iterated Local Search (ILS) based metaheuristic that defines an initial scheduling map aiming to minimize both application execution time and monetary cost. We point out that ILS~\cite{lourencco2003iterated} belongs to a class of local search algorithms proposed in the literature that efficiently investigates the solution space of combinatorial problems;
    \item  A  spot and on-demand burstable instance-based dynamic scheduler; 
    \item  Validation of Burst-HADS by presenting evaluation results from experiments conducted in  a real cloud provider environment, Amazon EC2, considering synthetic applications \cite{MELOALVES2017150} as well as NAS benchmark application in different scenarios subject to spot VM hibernations and resuming events.

\end{itemize}

Performance results from the experiments confirm the effectiveness of Burst-HADS in terms of monetary costs and execution times compared to an approach based only on Amazon EC2 on-demand VMs. Concerning the comparison with our previous work \cite{teylocc}, results show that the new proposed ISL-based primary scheduler and the use of burst instances can reduce the impact in the execution time of using hibernate-probe instances.

The remaining of this paper is organized as follows. Section~\ref{sec:related_work} discusses some related work. Section~\ref{sec:problem_definition} defines the system and application models, while Section~\ref{sub:hads} presents  the main modules of  Burst-HADS. Evaluation results from experiments on EC2 cloud are presented in Section~\ref{sec:experimental_tests}. Finally, Section~\ref{sec:conclusion} concludes the paper and introduces some future directions.

\section{Related Work}
\label{sec:related_work}    

Since AWS introduced the concept of burstable VM, many works exploring its features have been proposed. Many of them focus on evaluating the burstable approach and the improvement in computational performance that it can induce. In  \cite{leitner2015bursting},  Leitner and Scheuner present a first empirical and analytical study about the second generation of AWS burstable instances (T2 family). They specifically considered T2.micro, T2.small, and T2.medium instances. Their article aimed at answering if, in terms of monetary cost and performance, these instance types are more efficient than other instance ones. Results show that compared to general-purpose and computed-optimized instances (2015 generation), the evaluated T2 instances provide a higher  CPU performance-cost ratio as long as the average utilization of instances is below 40\%.


In order to figure out CPU usage limits on-the-fly, according to the dynamic variation of the workloads, Ali \textit{et al.} proposed  in  \cite{ali2018cedule} an autonomic framework that combines light-weight profiling and an analytical model  The objective is to maximize the amount of work done using the burstable capacity of T2 VM instances.  The authors state that the framework extends the CPU credits depletion period. Similarly to t Leitner and Scheuner's work \cite{leitner2015bursting}, their results also confirm the benefits from active CPU usage control when burstable instances are exploited. However, they do not discuss the impact in the total execution time when such an approach is applied. Jiang \textit{et al.} \cite{jiang2019burstable}  analytically modeled the performance of burstable VMs, considering their respective configuration, such as  CPU, memory, and CPU credits parameters.  They also show that providers can maximize their total revenue by finding the optimal prices for burstable instances.  Although their work, contrarily to ours,  does not focus on application performance,  its contribution is extremely interesting since it states that providers can offer burstable instances for low prices without losing revenue while meeting QoS parameters, including burst mode's performance.

Mehta and Chandy \cite{mehta2015leveraging} conducted a series of experiments in EC2 to evaluate the performance of a  job executing in instances T1.micro and T2.micro. The article shows it is possible to overcome  performance penalty due to lack of CPU credits by a simple checkpoint and migration approach. On the one hand, the latter is very similar to the Burst-HADS migration procedure. On the other hand, it differs from Burst-HADS and its previous versions \cite{teylocc} since it does not exploit spot VMs to minimize monetary cost. 

Some scheduling and scaling problems also take advantage of burstable instances features. In \cite{baarzi2019burscale}, for example, Baarzi \textit{et al.} propose an autoscale tool denoted BurScale, which uses burstable instances, together with on-demand instances, to handle transient queuing which arises due to traffic variability. They also present how burstable instances can mask VM startup/warmup costs when autoscaling, to handle flash crowds, takes place.  Using two distinct workloads, a stateless web server cluster and a stateful Memcached caching cluster, the authors show that a careful combination of burstable and regular instances ensured similar performance for applications as traditional autoscaling systems while reducing up to 50\% of the monetary cost.


In \cite{wang2017exploiting}, Wang \textit{et al.} combined on-demand, spot, and burstable instances proposing an in-memory distributed storage solution. Burstable instances are used as a backup to overcome performance degradation resulting from spot instance revocations. According to the authors, those instances' burst capacity makes them ideal candidates for such a backup. Performance results show that the backup that uses burstable instances presents a latency, which is 25\% lower than the latency of backup based on regular instances, inducing, therefore, significant monetary cost saving.

To deal with spot VM revocations, we proposed in \cite{teylo2019static} a static heuristic that creates pre-defined backup maps before the execution of the job tasks themselves. It was the first attempt to cope with spot VMs hibernation, and results from simulation showed that the hibernation problem is better handled with a dynamic approach. Thus, in \cite{teylocc} we present a dynamic scheduler, denoted HADS that uses both spot and regular (non-burstable) on-demand VMs to execute BoT applications. It aims at minimizing the execution's monetary cost respecting application deadline 
On the other hand, in the current work, Burst-HADS exploits hibernation-prone spot VMs and burstable VMs. It also applies new heuristics to reduce not only the monetary cost of the execution but also the execution time of the application, while meeting the deadline defined by the user.

To the best of our knowledge, the hibernation mechanism of the spot VMs is only discussed in our previous works \cite{teylo2019static, teylocc} and in Fabra \textit{et al.} \cite{fabra2019reducing}. In the latter, the authors consider a  scenario where hibernation-prone spot VMs can is used and then they show that deadline constraints add complexity to the problem of resource provisioning. However, no practical solution is presented neither discussed. Moreover, that work does not concern the task scheduling problem, but a resource provisioning one.

\section{Scheduling with Hibernation-prone Spot and Burstable On-demand Instances}

\subsection{System and Application Models}
\label{sec:problem_definition}

\begin{table}[H]
\centering
\caption{Notation used in the application and system models.\label{tab:samples}} 
\begin{tabularx}{\columnwidth}{c|X}
\toprule
\textbf{Name} & \textbf{Description} \\
\hline
            $B$ & Set of tasks\\
            $M$ & Set of VMs \\
            $M^s$ & Set of VMs spots\\
            $M^o$ & Set of VMs on-demand\\
            $M^b$ & Set of VMs burstable on-demand  VMs\\
            $T$   & Discretized time set\\ 
            
            $D$ &  Deadline defined by the user\\
            $D_{spot}$ &  Estimated time limit which ensures that there will be enough spare time to migrate tasks of a hibernated  spot VM to other VMs no matter when hibernations take place\\
            
            $\omega$        & Overhead of initialization of a VM  \\
            $vm_j$                     & Virtual machine \\ 
             $m_j$                      & Memory capacity of $vm_j$ in gigabytes\\ 
             $c_j$                      & Cost per period of time of $vm_j$ \\
             $vcpu_k$          & A Virtual Core of $vm_j$ \\                 
             $VC_j$                      & Set of vCPUs of $vm_j$ \\
             $cc_j$                     & CPU credits of $vm_j$ \\

             $t_i$                      & a task of the BoT application\\
             $rm_i$                     & Amount of memory required by a task $t_i$ \\ 
             $e_{ij}$                   & Time required to execute task $t_i$ in a $vm_j$\\ 
\bottomrule
\end{tabularx}
\end{table}

Let $M = M^s \cup M^o \cup M^b$ be the user-provided set of VMs that can be used to execute the BoT application, where $M^{s} \subset M$ is the set of spot instances, $M^{o} \subset M$ is the set of non burstable on-demand ones and $M^{b} \subset M$ is the set of burstable on-demand VMs.   Let $D$ be the application deadline, defined by the user, and $T = \{1, \ldots, D\}$ be a discrete set  of time periods.  Each $vm_j \in M$ has a memory capacity of $m_j$ gigabytes, $|VC_j|$ virtual cores (vCPU), and a current CPU credit amount $cc_j$ (in the case of $vm_j \not\in M^{b}$, i.e., non-burstable VMs, $cc_j=\infty$). 

By default, Amazon  EC2 does not allow more than five VMs of similar types and markets to run simultaneously. Thus,  if the user decides, for example, to use only on-demand VMs of type $C5.xlarge$, spot VMs of type $C5.2xlarge$, and burstable VMs of type $T3.xlarge$,  there will be   a maximum  of  five VMs of each type at each set  $M^{o}$, $M^{s}$ and $M^{b}$.

When the user requests a new VM, there exists a time delay for using it which due to the launch of the VM by the provider plus its operating system booting.
Therefore, in this work, we consider the same overhead initialization time cost $\omega$ for all VMs, which comprises these waiting times. When a new $vm_j$ is launched, after $\omega$ periods of time, the user is charged $c_j$ for each period of time. When the VM terminates, the user's charge for this VM immediately stops.  

Let $B$ be the set of tasks of a BoT application. We assume that each task $t_i \in B$ requires a known amount of memory $rm_i$, and it is executed in only one $vcpu_k \in VC_j$ of $vm_j$. A multi-core $vm_j$ ($|VC_j| > 1$) can execute more than one task simultaneously (one task per core) only when there is enough main memory to allocate them in the VM. We also consider that the execution time $e_{ij}$ of each task $t_i$ in any $vm_j \in M$ is known beforehand. In the case of a burstable $vm_j \in M^b$,  $e_{ij}$ is the execution time of task $t_i$ in $vm_j$ in burst mode (100\% of the $vm_j$ processing power).

All variables and parameters defined in this section are summarized in Table \ref{tab:samples}.

\subsection{Burst Hibernation-Aware Dynamic Scheduler Framework}
\label{sub:hads}

The Burst Hibernation-Aware Dynamic Scheduler (Burst-HADS) that we propose has two main scheduling modules: i) the \textbf{Primary Scheduling Module}, that defines an initial scheduling map; and ii) a \textbf{Dynamic Scheduling Module} responsible for migrating tasks or executing task rollback recovering in idle VMs. Note that each of the modules can be easily changed or extended adopting a new approach. For instance, a new heuristic or metaheuristic can be included in the Primary Scheduling Module. The source code of Burst-HADS is available at \url{https://github.com/luanteylo/Burst-HADS}.

\subsection{Primary Scheduling Module}

It is worth remembering that the problem of scheduling tasks to distributed computing resources is an NP-complete one~\cite{ullman1975np}, even in simple scenarios. Furthermore, some features of clouds render it more difficult. We thus model the primary scheduling problem as a multi-objective integer programming problem whose objectives are to minimize both the monetary cost and the total execution time of the application. In other words, in our case, it is defined as the problem of scheduling tasks to VMs respecting the available storage capacity and trying to minimize the makespan and the monetary costs. 

Aiming at ensuring the application deadline $D$  no matter if and when spot VM hibernations could happen,  in our previous work  \cite{teylocc}, we defined $D_{spot}$  as the worst-case estimated makespan, which guarantees that there will always have enough spare time to migrate tasks of any hibernated spot VM to other VMs and execute them. $D_{spot}$ is computed by considering the longest tasks that might need to be migrated and executed to/in the system's slowest virtual machines.

Let the binary variable $X_{ijk}^v$ indicate whether a {\em task} $t_i \in B $ allocated to a  $vcpu_k \in VC_j$ of a  $vm_j \in M^s$  will start executing ($X_{ijk}^v=1$), or not ($X_{ijk}^v=0$), at time period $v \in T$. Let also $Z_j$ and $ZT$ be variables which respectively keep the last period of execution  of a $vm_j \in M^s$ and the total execution time of the application (makespan).

The  proposed objective function (Equation \ref{eq:obj}) is a  weighted function that minimizes   the monetary cost  and the  makespan,  where $\alpha$  and $1-\alpha$ are the respective weights given by the user for the objectives.   Note that, in this case, both the monetary cost and the makespan  have to be first normalized. The objective function  is subject to the constraints given by equations \ref{eq:memory}, \ref{eq:vcpu}, \ref{eq:aloc}, \ref{eq:deadline}, and \ref{eq:makespan}. 

\begin{equation}
    \begin{split}
        \min{(\alpha\sum_{vm_j \in M^s}{Z_jc_j} + (1 - \alpha)ZT)}
    \end{split}
    \label{eq:obj}
\end{equation}

Subject to: 

\begin{equation}
    \begin{split}
        & \sum_{t_i \in B}\sum_{vcpu_k \in VC_j}{\sum_{q=p}^{v}{rm_i X_{ijk}^q }} \leq m_j, \\
        & \forall vm_j \in M^s,  \forall v \in T, \text{ and $p=\max(v - e_{ij}, 1)$}
    \end{split}
    \label{eq:memory}
\end{equation}

\begin{equation}
    \begin{split}
        & \sum_{t_i \in B}\sum_{vcpu_k \in VC_j}{\sum_{q=p}^{v}{X_{ijk}^q }} \leq |VC_j|,  \\
        & \forall vm_j \in M^s, \forall v \in T, \text{ and $p=\max(v - e_{ij}, 1)$}
    \end{split}
    \label{eq:vcpu}
\end{equation}

\begin{equation}
    \begin{split}
        & \sum_{vm_j \in M^s}\sum_{vcpu_k \in VC_j}{\sum_{v \in T}{X_{ijk}^v }} = 1, \forall i \in B
    \end{split}
    \label{eq:aloc}
\end{equation}

\begin{equation}
    \begin{split}
        & X_{ijk}^v  (v + e_{ij}) \leq Z_j  \\
        & Z_j \leq D_{spot} \\
        & \forall t_i \in B, \forall vm_j \in M^s, \forall vcpu_k \in VC_j \text{ and } \forall v \in T 
    \end{split}
    \label{eq:deadline}
\end{equation}

\begin{equation}
    \begin{split}
        & X_{ijk}^v Z_j \leq ZT  \\
        & \forall t_i \in B, \forall vm_j \in M^s, \forall vcpu_k \in VC_j \text{ and } \forall v \in T 
    \end{split}
    \label{eq:makespan}
\end{equation}

In Equations \ref{eq:memory} and \ref{eq:vcpu}, we define the constraints in terms of  memory and number of virtual cores, respectively.  Equation \ref{eq:aloc} determines that every task $t_i \in B$ must be allocated and started only once, in a a single $vcpu_k$ of  $vm_j \in M^s$  ($vcpu_k \in VC_j$). The constraint of Equation \ref{eq:deadline} ensures that $D_{spot}$ is respected while equation \ref{eq:makespan} defines the makespan constraint. 

In order to create the initial scheduling map, our previous Primary Scheduling Module, HADS, proposed in \cite{teylocc}, executes a simple greedy heuristic, which only minimizes monetary cost while respecting the application deadline. In the current work, we propose an Iterated Local Search (ILS) whose goal is to minimize both the monetary cost and the execution time of that initial scheduling map. ILS~\cite{lourencco2003iterated} is a metaheuristic that aims at improving a final solution by sampling in a broader and distant neighborhood of candidate solutions, and then applying a local search technique to refine solutions to their local optima. It explores a sequence of solutions created by perturbations of the current best solution to reach these distant neighborhoods.

After finding a scheduling map with ILS, a second heuristic is applied to include burstable instances into the solution. In case of spot VM hibernations, these instances will be used, in burst mode, by the dynamic scheduler, as an attempt to minimize the impact of these hibernations in the monetary cost and/or the execution time. 
Therefore, the Primary Task Scheduling algorithm (Algorithm \ref{alg:ils}) has two parts:  i) the iterated local search, and ii)  the burstable instances allocation. Furthermore, in line \ref{ils:greedy}, an initial solution  is generated  by calling Algorithm \ref{alg:initial_solution},  proposed, originally in \cite{teylocc}. The latter is a greedy heuristic that schedules the set of tasks $B$ to a set $M^s$ of spot VMs.

Initially,  Algorithm \ref{alg:initial_solution} sorts the tasks of $B$ in descending order by their memory size requirements (line \ref{initial_s:sort}). Calling the procedure \textit{check\_schedule}, it then verifies if it is possible to schedule task $t_i$ in the selected  $vm_j$ of the current phase, i.e., the task is scheduled to the VM if the memory requirements are satisfied and if  $D_{spot}$ is not violated. Note that scheduling tasks in an already allocated VM avoid VM deploying time. Thus, for each task $t_i \in B$, the algorithm tries to schedule it to a core of a virtual machine $vm_j$ from $A$, the set of already selected VMs (lines \ref{algP:start_ph1} to \ref{algP:end_ph1}). If task $t_i$ can not be scheduled to a  $vm_j$ of $A$, the  algorithm tries to select a new spot VM (lines \ref{algP:start_ph2} to \ref{algP:end_ph2}) using a weighted round-robin algorithm (WRR) \cite{katevenis1991weighted}. In  WRR, each spot VM has an associated weight, and the algorithm selects the VMs in a round-robin way, according to such weights. As shown in Equation \ref{eq:weight}, the weight of  $vm_j$, $weight(j)$, is equal to the quotient between $Gflops_j$ of $vm_j$, and $c_j$, the price of the VM per time period.  The $Gflops_j$ of  $vm_j$ is estimated using the LINPACK benchmark \cite{dongarra2003linpack} and expresses the computing power of this VM. Our choice in using WRR and spot VMs with different configurations is in agreement with Amazon's recommendations\footnote{\url{https://aws.amazon.com/pt/ec2/spot/instance-advisor/}} that say that an application should use different types of spot VMs to increase the availability of spot VM instances. According to Kumar et al. \cite{kumar2018survey}, interruptions of spot VMs, which include hibernation, usually take place in VMs of the same type. Therefore, a choice of heterogeneous spot VMs minimizes the impact of possible VM hibernations. Finally, in line \ref{AlgP:createS},  Algorithm \ref{alg:initial_solution} calls the function $create\_solution$ that receives the set of the selected VMs $A$ and returns a solution $S$ that defines a scheduling map.

\begin{equation}
    weight(vm_j) =  Gflops_j/ c_j, \textit{where $vm_j \in M$}
    \label{eq:weight}
\end{equation}

\begin{algorithm}[htb]
		\caption{\textit{Primary Task Scheduling}}
		\label{alg:ils}
		\small
        \begin{algorithmic}[1]
		\renewcommand{\algorithmicrequire}{\textbf{Input:}}
		\renewcommand{\algorithmicensure}{\textbf{Output:}}
        \REQUIRE $B$, $M$, $M^s$, $M^o$, $M^b$, $max\_ondemand$, $max\_iterations$, $max\_failed$, $relaxed\_rate$, $D_{spot}$ \textbf{ and }  $D$

        \STATE \COMMENT{/*PART 01 - Iterated Local Search*/}
        
        \STATE $S \leftarrow initial\_solution(B, M^s, D_{spot})$ \label{ils:greedy} \COMMENT{/*Algorithm \ref{alg:initial_solution}*/}

        \STATE $S \leftarrow local\_search(S, max\_attempt, swap\_rate)$ \label{ils:localsearch1} \COMMENT{/* Algorithm \ref{alg:localsearch}*/} 
        \STATE $S_{best} \leftarrow S$ 
        \STATE $RD_{spot} \leftarrow D_{spot}$

        \STATE $i \leftarrow 0$ \label{ils:part2_start}
        \STATE $last\_best \leftarrow i$
        \WHILE{$i < max\_iteration$}
        
            \STATE \COMMENT{/* Perturbation */} 
            \STATE $vm_j \leftarrow random\_choice(M^s)$ \label{ils:perturbation1:start}
            \STATE $S.selected\_vms \leftarrow S.selected\_vms \cup vm_j$
            \STATE $M^s \leftarrow M^s \backslash \{vm_j\}$ \label{ils:perturbation1:end}
            
            \STATE $failed \leftarrow i - last\_best$ \label{ils:perturbation2:start}
            \IF{$failed > max\_failed$} 
                \STATE $RD_{spot} \leftarrow RD_{spot} + (relax\_rate \times RD_{spot})$
            \ENDIF  \label{ils:perturbation2:end} 
            \STATE $S \leftarrow local\_search(S,max\_attempt, swap\_rate)$ \label{ils:localsearch2}
             
            \IF{$fitness(S, D_{spot}) < fitness(S_{best}, D_{spot})$} 
               \STATE $S_{best} \leftarrow S$
               \STATE $last\_best \leftarrow i$
            \ENDIF
        
            \STATE $i \leftarrow i + 1$
            
        \ENDWHILE \label{ils:part1_end}
    
    \STATE \COMMENT{/*PART 02: Burstable instance allocation*/} \label{ils:part3:start}
    
    \STATE $n \leftarrow \lceil burst\_rate \times S.selected\_vms \rceil$  \label{ils:burst_rate}
    \STATE $S_{final} \leftarrow burst\_allocation(S_{best}, burst\_rate, M^b, D_{spot}, D)$  \label{ils:part3:end}
    
    \STATE $create\_primary\_map(S_{final})$\label{ils:map} 
    
	\end{algorithmic}
\end{algorithm}

\begin{algorithm}[htb]
        \caption{\textit{Initial Solution}}
        \label{alg:initial_solution}
        \small
         \begin{algorithmic}[1]
        \renewcommand{\algorithmicrequire}{\textbf{Input:}}
		\renewcommand{\algorithmicensure}{\textbf{Output:}}
		\REQUIRE $B$, $M^s$, $D_{spot}$
		\STATE $sort\_by\_memory(B)$ \label{initial_s:sort} \COMMENT{\footnotesize  /*Sort Tasks by memory requirement $rm_i$*/}
        \STATE $A \leftarrow \emptyset $ \COMMENT{\footnotesize  /*Set of selected VMs*/}
        
        \FOR{$ \textbf{ all } t_i \in B$} \label{algP:mainfor}
          
          \STATE \COMMENT{\footnotesize (/*Phase 1: Try to schedule the task in an already selected spot VM*/}
          \STATE $sort\_by\_price(A)$ \COMMENT{\footnotesize/*Sort the selected VMs by price*/}
          \FOR{\textbf{ all }  $vm_j \in A$}\label{algP:start_ph1}
	                \IF{$check\_schedule(t_i, vm_j, D_{spot})$}
     			        \STATE $schedule(t_i, vm_j)$ 
     			        \STATE $break$ \COMMENT{\footnotesize /*Schedule next task*/}
     			     \ENDIF
        \ENDFOR\label{algP:end_ph1}
        
        \STATE \COMMENT{/*Phase 2: Try to schedule the task in a new spot VM*/} 
 			\IF{$\textbf{ not } scheduled$} \label{algP:start_ph2}
  			     \STATE $vm_k \leftarrow get\_wrr\_VM(M^s)$\label{algP:rr}\COMMENT{\footnotesize /*Select a spot VM using the weighted round-robin heuristic*/}   
      			 \IF{$check\_schedule(t_i, vm_k, D_{spot})$}
     			    \STATE $schedule(t_i, vm_k)$
      			    \STATE $A \leftarrow A \cup \{vm_k\}$ \COMMENT{/*Update the set of selected VMs*/} \label{AlgP:update1}
      			    \STATE $M^s \leftarrow M^s \backslash \{vm_k\}$
      			    \STATE $break$ \COMMENT{/*Schedule next task*/}
  		        \ENDIF 
 	        \ENDIF \label{algP:end_ph2}

        \ENDFOR \label{algP:mainfor_end}
        
    \STATE $S \leftarrow create\_solution(A)$ \label{AlgP:createS}
	\STATE \textbf{Return} $S$ \COMMENT{*/Returns a scheduling solution $S$*/}
         \end{algorithmic}
\end{algorithm}
 
After obtaining an initial solution, the ILS algorithm  tries to improve it by applying local search and perturbation procedures (Algorithm \ref{alg:ils}, lines \ref{ils:localsearch1} - \ref{ils:part1_end}). Thus, let $S$  be a solution that defines a scheduling map of all tasks $t_i \in B$ to a subset of VMs of $M$. Let $fitness(S)$ be a weighted function that assigns a value to the quality of $S$. Since this function is equivalent to the objective function presented in Equation \ref{eq:obj}, we define in Equation \ref{eq:fitness} the $fitness(S, D_{spot})$ function, where $cost$ is the total monetary cost of $S$ and $mkp$ is the total execution time of the application.

\begin{equation}
\label{eq:fitness}
fitness(S, D_{spot})= \begin{cases}

    \infty, {\footnotesize \text{if violates $D_{spot}$}}\\
    \alpha.cost + (1- \alpha).mkp, {\footnotesize \textit{otherwise}}
\end{cases}
\end{equation}

 Algorithm~\ref{alg:ils} executes a local search by calling, in line~\ref{ils:localsearch1}, the $local\_seach$ procedure of Algorithm~\ref{alg:localsearch}, which executes a series of attempts to improve the current solution by swapping tasks between the selected VMs. Algorithm~\ref{alg:localsearch} receives as input the current solution $S$, the $max\_attempt$ parameter that determines the number of times the local search will be executed, the set $B$  and the $swap\_rate \in ]0, 1]$ value that determines the number of tasks that will be swapped at each iteration. The $swap\_rate$ parameter is tuned before the execution. All the parameters used in our tests, including the $swap\_rate$, $max\_attempt$, $max\_iteration$ and other parameters will be presented in Section \ref{sec:experimental_tests}.
 
 As can be observed in lines \ref{localseach:dest} and \ref{localsearch:array} of the Local Search algorithm (Algorithm~\ref{alg:localsearch}), a solution $S$ is composed of two structures: (i) a vector, which controls task allocation, where indexes correspond to tasks, and each element keeps the identity of the VM that will execute the corresponding task, and (ii) a list composed by selected VMs of $M$. Firstly, the algorithm computes the number of tasks that will be swapped at each iteration (line \ref{localsearch:ntasks}) and randomly selects a destination VM ($vm_{dest}$, line \ref{localseach:dest}). After that, the algorithm starts the  tasks swapping procedure (lines \ref{localsearch:searchloop_start} to \ref{localsearch:searchloop_end}), where  $n$ tasks,  also randomly selected (line \ref{localseach:task_selection}), are moved to the $vm_{dest}$. After each swap $movement$, the $local\_seach$ procedure checks if the quality of the new generated solution has improved (line \ref{localsearch:check}) and it updates the $S_{best}$ solution, if necessary. In the end, the procedure returns the best-found solution (line \ref{localseach:return}).

{
\begin{algorithm}[htb]
		\caption{\textit{Local Search}}
		\label{alg:localsearch}
		\footnotesize
        \begin{algorithmic}[1]
		\renewcommand{\algorithmicrequire}{\textbf{Input:}}
		\renewcommand{\algorithmicensure}{\textbf{Output:}}
        \REQUIRE $S$, $max\_attempt$, $swap\_rate$ \textbf{ and } $D_{spot}$
        
        \STATE $S_{best} \leftarrow S$
        \STATE $n \leftarrow swap\_rate \times |B|$ \label{localsearch:ntasks} 
        \STATE $attempt \leftarrow 0$

        \STATE $vm_{dest} \leftarrow random\_choice(S.selected\_vms)$  \label{localseach:dest}
        
        \WHILE{$attempt < max\_attempt$} \label{localsearch:searchloop_start}
            \FOR{$k \in \{1,..., n\}$}
                \STATE $t_i \leftarrow random\_choice(B)$ \label{localseach:task_selection}
                \STATE $S.allocation\_array[t_i] = vm_{dest}$ \label{localsearch:array}
                \IF{$fitness(S, D_{spot}) < fitness(S_{best}, D_{spot})$} \label{localsearch:check}
                    \STATE $S_{best} \leftarrow S$
                \ENDIF
            \ENDFOR
            
            \STATE $attempt \leftarrow attempt + 1$
        \ENDWHILE  \label{localsearch:searchloop_end}
        
        \STATE \textbf{return } $S_{best}$ \label{localseach:return}
	\end{algorithmic}
\end{algorithm}
 }  

 After the first execution of the $local\_search$ procedure (line \ref{ils:localsearch1}), Algorithm \ref{alg:ils} has a loop that firstly performs a  perturbation (lines \ref{ils:perturbation1:start}-\ref{ils:perturbation2:end}) and  then a new local search (line \ref{ils:localsearch2}). The perturbation is responsible for diverting the metaheuristic from local optimal solutions. In the current work, we use two simple perturbation strategies. The first one includes a not selected spot $vm_j \in M^s$ into the current solution $S$ (lines \ref{ils:perturbation1:start}-\ref{ils:perturbation1:end}). The second one, called relaxing perturbation,  increases the $D_{spot}$ limit (lines \ref{ils:perturbation2:start}-\ref{ils:perturbation2:end}). Note that the latter is executed only when the number of iterations without finding a better solution (called $failed$ iterations) is higher than the $max\_failed$ parameter.  In this case, the metaheuristic increases the $D_{spot}$ limit in $relax\_rate$ percent, where $relax\_rate \in ]0, 1]$ is also a parameter defined by the user.

Upon finishing the ILS (Part 1), Algorithm~\ref{alg:ils} executes the $burst\_allocation$ procedure (Part 2, lines \ref{ils:part3:start}-\ref{ils:part3:end}). A number of burstable VMs,   which is a percentage of the number of used spot VMs, $burst\_rate$,   are included in the solution.  Since the relaxed perturbation leads some tasks to violate the  $D_{spot}$ limit, the $burst\_allocation$ procedure moves these tasks to the burstable instances. Each of them can receive at most one task to be executed in baseline mode.  However, if there still exist tasks violating $D_{spot}$ and no available burstable VM, the procedure allocates them to the cheapest regular on-demand  VMs. On the other hand,  if a  burstable VM  remains idle, the task with the latest finishing time in the scheduling map is moved to it.  
  
Remark that our strategy of having a single task per burstable instance at a time, executing in baseline mode, induces CPU credits accumulation. Consequently, these burstable instances became the best candidates to receive tasks in case of hibernations.

\subsection{Dynamic Scheduling Module}
\label{sec:dynamic_module}

The Dynamic Scheduling Module performs some actions in response to events, such as spot VM hibernation, resuming, and  idleness, that may occur along with the execution. Let \textit{BR}, \textit{IR}, \textit{HR}, and \textit{TR} of $M$ be the set of \textit{busy}, \textit{idle}, \textit{hibernated}, and \textit{terminated} VMs respectively. Thus, Burst-HADS considers that a $vm_j$ can be in one of the following states i) \textit{busy}, if active and executing tasks ($vm_j \in BR$); ii) \textit{idle}, if active but not executing any task ($vm_j \in IR$); iii) \textit{hibernated}, if it has been hibernated by the cloud provider ($vm_j \in HR$); and iv) \textit{terminated}, if the VM has terminated or it was not available at the beginning of the application execution ($vm_j \in TR$). 

One of the main goals of the Dynamic Scheduling Module is to decide when a VM should terminate. On the one hand, as  VMs are charged for each period of time (see Section \ref{sec:problem_definition}), the user has an interest that a VM terminates as soon it becomes \textit{idle}, aiming for cost reduction and avoiding extra ones. On the other hand, in some cases, it might be interesting to keep it because it can receive and execute tasks without incurring deploying overhead.  Hence, Burst-HADS has a termination policy where the allocation time is logically divided into units denoted Allocation Cycles (ACs). A $vm_j$ that reaches the end of its current AC, denoted $AC\_cur_j$, in \textit{idle} state, is terminated. 

The Dynamic Scheduling Module handles the following events related to a VM, $vm_j$: i) $vm_j$ becomes idle; ii) $vm_j$, which is idle, reaches the end of $AC\_cur_j$; iii) $vm_j$ is a spot VM that hibernates; and iv) $vm_j$, which has hibernated, resumes. 
The following actions are taken for each of the events:
\begin{itemize}
\item{\it $vm_j$ becomes \textit{idle}}:
When a $vm_j$ finishes the execution of all tasks scheduled to it, its state is updated to \textit{idle} and, if there exist scheduled tasks that have not completely executed, a work-stealing procedure starts at the beginning of the next $vm_j$'s $AC$, assigning some or all of these tasks to $vm_j$, whose state is, thus, set to  \textit{busy}.

\item{\it Non-burstable idle $vm_j$ reaches the end of $AC\_cur_j$}:

If at the end of $AC\_cur_j$ a non-burstable $vm_j$ remains \textit{idle}, it is removed from the set   IR of \textit{idle} VMs and included into $TR$, the set of \textit{terminated} VMs. 

\item{\it  Spot $vm_j$ hibernates}:
 The $vm_j$'s state is updated to \textit{hibernated} and if it was  $busy$, Burst-HADS  starts the  \textit{burst migration procedure} (see Section \ref{sec:migration}).

\item{\it Spot $vm_j$ resumes}:
When a hibernated spot $vm_j$ resumes, it is excluded from  $HR$, the set of hibernated VMs, and Burst-HADS starts a work-stealing procedure that tries to move tasks from $busy$ VMs to $vm_j$ (see Section \ref{sec:work_stealing}).
\end{itemize}

\subsection{Burst Migration Procedure}
\label{sec:migration}

As previously explained, as soon as a spot $vm_l$ hibernates, Burst-HADS executes the migration procedure, searching for a set of VMs to assign and execute non-finished tasks, denoted affected tasks, that were previously scheduled to $vm_l$. Algorithm \ref{alg:migration} presents the migration procedure, which always respects the deadline $D$ when selecting tasks to migrate. It receives as input the set $Q_l \subset B$ of affected tasks, the sets of idle, busy, and non-launched regular on-demand VMs ($IR$, $BR$  and $M^o$, respectively), and the deadline $D$.

{
\begin{algorithm}[htb]
		\caption{\textit{Burst Migration Procedure}  \label{alg:migration}}
		\footnotesize
\begin{algorithmic}[1]
      
		\renewcommand{\algorithmicrequire}{\textbf{Input:}}
		\renewcommand{\algorithmicensure}{\textbf{Output:}}
        \REQUIRE  $Q_l$, $IR$, $BR$, $M^o$, \textbf{ and } $D$
        
        \STATE $Q_l \leftarrow  sort\_tasks(Q_l)$  \COMMENT{/* Prioritizes tasks with checkpoints */} \label{alg:migration_proc:sort_tasks}
       
        \FOR{\textbf{each} $t_i \in Q_l$}
        
          \STATE \COMMENT{/*Attempt 1 - Try to migrate task to a Burstable IDLE VM */}
            
            \FOR{\textbf{each burstable } $vm_j \in  IR$} \label{alg:migration_proc:at1_start}
               
                \STATE $rcc_{ij} \leftarrow \lceil e_{ij} / burst\_period \rceil$ \label{alg:migration_proc:cpu_credits}
            
                \IF{$cc_j > rcc_{ij} \textbf{ and } check\_migration(t_i,vm_j, D)$} \label{alg:migration_proc:check_at1}

                    \STATE $reserve\_cpu\_credits(rcc_{ij}, vm_j)$  \label{alg:migration_proc:reserve}
                    \STATE $set\_burst\_mode(t_i, vm_j)$\label{alg:migration_proc:set_burst}
                    \STATE $migrate(t_i, vm_j)$  \label{alg:migration_proc:migrate_burst}
                    \STATE $IR \leftarrow IR \backslash \{vm_j\}$  \label{alg:migration_proc:idle_burst1}
                    \STATE $BR \leftarrow BR \cup \{vm_j\}$ \label{alg:migration_proc:idle_burst2}
                    \STATE $break$ \COMMENT{/*Migrate next task*/} 
                \ENDIF
            \ENDFOR  \label{alg:migration_proc:at1_end}
       
            \IF{\textbf{ not }$migrated$}
              
                \STATE \COMMENT{/*Attempt 2 - Try to migrate task to a NON-burstable Idle VM*/}
                \STATE $sort\_by\_market(IR)$ \label{alg:migration_proc:sort1} \COMMENT{/* Prioritizes spot VMs */}
                \FOR{\textbf{each NON-burstable} $vm_j \in IR$} \label{alg:migration_proc:at2_start}
                  \IF{$check\_migration(t_i,vm_j, D)$} \label{alg:migration_proc:at2_check}
                        \STATE $migrate(t_i, vm_j)$ 
                        \STATE $IR \leftarrow IR \backslash \{vm_j\}$ 
                        \STATE $BR \leftarrow BR \cup \{vm_j\}$
                        \STATE $break$ \COMMENT{/*Migrate next task*/} 
                    \ENDIF
                  
                \ENDFOR  \label{alg:migration_proc:at2_end}
            \ENDIF
              \IF{\textbf{ not }$migrated$}
                \STATE \COMMENT{/*Attempt 3 - Try to migrate task to a NON-burstable Busy VM*/}
                \STATE $sort\_by\_market(BR)$ \label{alg:migration_proc:sort2} \COMMENT{/* Prioritizes spot VMs */}
                \FOR{\textbf{each NON-burstable} $vm_j \in BR$} \label{alg:migration_proc:at3_start}
                    \IF{$check\_migration(t_i, vm_j,  D) $}  \label{alg:migration_proc:at3_check}
                        \STATE $migrate(t_i, vm_j)$
                        \STATE $break$ \COMMENT{/*Migrate next task*/}
                    \ENDIF
                \ENDFOR  \label{alg:migration_proc:at3_end}
            \ENDIF
            
            \IF{\textbf{ not }$migrated$}
                \STATE \COMMENT{/*Attempt 4 - Try to migrate task to a new NON-burstable on-demand VM*/}
                \STATE $sort\_by\_price(M^o)$
                \FOR{\textbf{each} $vm_j \in M^o$} \label{alg:migration_proc:at4_start}
                    \IF{$start_{ij} + e_{ij} + \alpha < D$} \label{alg:migration_proc:at4_check}
                        \STATE $start\_vm(vm_j)$ 
                        \STATE $migrate(t_i, vm_j)$ 
                        \STATE $M^o \leftarrow M^o \backslash \{vm_j\}$ \label{alg:migration_proc:at4_update1}
                        \STATE $BR \leftarrow BR \cup \{vm_j\}$              \label{alg:migration_proc:at4_update2}
                        \STATE $break$ \COMMENT{/*Migrate next task*/}
                    \ENDIF
                \ENDFOR  \label{alg:migration_proc:at4_end}
            \ENDIF
        \ENDFOR
         
	\end{algorithmic}
\end{algorithm}
 }   

Initially, $Q_l$ is ordered, giving priority to those tasks that were executing at the moment of the hibernation and had been previously checkpointed (line \ref{alg:migration_proc:sort_tasks}). Burst-HADS applies the checkpoint/recovery mechanism that we proposed and evaluated in our previous work \cite{teylo2020developing} where checkpoints are taken with the help of Checkpoint Restore In Userspace (CRIU) \cite{criu},  a widely used tool that records the state of individual applications.
In order to avoid the overhead of launching new VMs, the migration procedure  gives priority to the use of already launched VMs. For each task $t_i \in Q_l$ the algorithm first tries to migrate the task to an \textit{idle} burstable VM (lines \ref{alg:migration_proc:at1_start} to \ref{alg:migration_proc:at1_end}). Otherwise, it tries to schedule $t_i$ to one of the non-burstable \textit{idle} VMs of set $IR$ (lines \ref{alg:migration_proc:at2_start} to \ref{alg:migration_proc:at2_end}). If not possible, it tries to schedule $t_i$ to a non-burstable \textit{busy} VM of set $BR$ (lines \ref{alg:migration_proc:at3_start} to  \ref{alg:migration_proc:at3_end}).  

Tasks migrated to burstable VMs are executed in burst mode, i.e., using 100\% of the VM's CPU processing power. Therefore, it is necessary to guarantee that a selected burstable VM will have enough CPU credits to execute all tasks assigned to it. Let $burst\_period$ be the number of periods of $T$ corresponding to one credit consumption in burst mode. Since we consider that a task $t_i$ is executed in only one core (see Section \ref{sec:problem_definition}), it is possible to estimate (line \ref{alg:migration_proc:cpu_credits})  the number of CPU credits consumed  by task $t_i$, where $e_{ij}$ is the execution time of $t_i$ on $vm_j$ in burst mode,  and $rcc_{ij}$ is the estimated number of required CPU credits. Then, the algorithm checks if $vm_j$ has enough CPU credits and call the $check\_migration$ function to guarantee that $vm_j$ has enough memory and will execute $t_i$ before the deadline $D$  (line \ref{alg:migration_proc:check_at1}).   If both conditions are satisfied, the $rcc_{ij}$ CPU credits are reserved for $t_i$ (line \ref{alg:migration_proc:reserve}),  the burst mode is set up (line \ref{alg:migration_proc:set_burst}), and $t_i$ is migrated to $vm_j$ (line \ref{alg:migration_proc:migrate_burst}). In this case the burstable $vm_j$ is removed from set $IR$ and included in set $BR$ (lines \ref{alg:migration_proc:idle_burst1} and \ref{alg:migration_proc:idle_burst2}). Note that the cloud provider continuously updates  the number of credits, $cc_j$, of every burstable VM.

Moreover, when Algorithm \ref{alg:migration} migrates a task $t_i$  to a non-burstable \textit{idle} $vm_j$,  it also calls the function $check\_migration$ to verify if the VM has enough memory and will be able to finish the task before the deadline $D$ (lines \ref{alg:migration_proc:at2_check}  and \ref{alg:migration_proc:at3_check}). Furthermore,  if $vm_j$ is a spot VM, the function $check\_migration$ should also verify if there will be enough spare time in $vm_j$ between the end of the execution of $vm_j$ tasks (including task $t_i$) and the deadline $D$ since, in this case, 
$vm_j$ is also subject to hibernation.  The spare time has to be greater than the execution time of the longest task scheduled to $vm_j$,  
ensuring, therefore, that if a hibernation occurs, there will be enough time to migrate and execute all affected tasks before the deadline $D$.

Finally, if there does not exist any available  already deployed VM able to execute task $t_i$, the algorithm migrates the task to a new  on-demand VM of set $M^o$ (lines \ref{alg:migration_proc:at4_start} to \ref{alg:migration_proc:at4_end}). In this case, it is necessary to verify that, considering the start period of $t_i$ in $vm_j$ ($start_{ij}$), its execution time ($e_{ij}$) plus the overhead $\alpha$ will not violate the deadline  (line \ref{alg:migration_proc:at4_check}). The new allocated on-demand VM is then removed from set $M^o$ and included in set $BR$ (lines \ref{alg:migration_proc:at4_update1} and \ref{alg:migration_proc:at4_update2}).

\subsection{Burst Work-Stealing} 
\label{sec:work_stealing}

The work-stealing procedure, presented in Algorithm \ref{alg:work_steal}, aims at reducing the allocation time of regular on-demand VMs as well as balancing the load of spot VMs. It is triggered  when a hibernated spot VM resumes or when a VM (spot or on-demand)  becomes {\em idle}. Basically, the procedure tries to move tasks from non-burstable  \textit{busy} VMs to the \textit{idle} VM.

For each non-burstable \textit{busy} $vm_j \in BR$ the procedure selects the tasks that can be stolen from it (line \ref{alg:work_steal:select}) and tries to migrate them to the \textit{idle} $vm_k$ (lines \ref{alg:work_steal:alloc:start}-\ref{alg:work_steal:alloc:end}).  Since regular on-demand VMs are more expensive than spot ones, the procedure considers firstly the tasks from the former (line \ref{alg:ws:sort1}).

Similarly to the migration procedure, for each selected task of a $vm_j$, the work-stealing procedure also verifies, by calling the function $check\_migration$, if the task migration would result in the  deadline violation (line \ref{alg:work_steal:alloc:check}). Since tasks migrated to burstable VMs by the work-stealing procedure are executed in the baseline mode, after verifying if the \textit{idle} $vm_k$ is burstable (line \ref{alg:work_stel:isburst}), the algorithm sets up the execution to the baseline mode (line \ref{alg:migration_proc:set_base}). In this work, to set up a burstable VM to the baseline mode means that the task cannot run using 100\% of the CPU processing power, but uses only the baseline performance defined by the provider\footnote{\url{https://docs.aws.amazon.com/AWSEC2/latest/WindowsGuide/burstable-credits-baseline-concepts.html}}. To limit the CPU utilization, Burst-HADS uses the cpulimit tool \cite{cpulimit}, an open-source program that limits the CPU usage of a process in GNU/Linux.   Moreover, if the \textit{idle} $vm_k$ is burstable, only one task is moved to it, to avoid  a queue of tasks to be executed at the baseline mode,  which could increase the application total execution time. Thus, after moving one task to the burstable $vm_k$ (line \ref{alg:work_steal:stop}), the algorithm stops the loop.
Finally, if at least one task is migrated to $vm_k$, its state changes to \textit{busy} and, consequently, it is included into the set of \textit{busy} VMs and removed from the set of \textit{idle} ones (lines \ref{alg:work_steal:alloc:busy} and  \ref{alg:work_steal:remove:idle}).

{
\begin{algorithm}[H]
		\caption{\textit{Burst Work-Stealing Procedure}        \label{alg:work_steal}}
		\footnotesize
        \begin{algorithmic}[1]
 
		\renewcommand{\algorithmicrequire}{\textbf{Input:}}
		\renewcommand{\algorithmicensure}{\textbf{Output:}}
        \REQUIRE $BR$, $IR$, $vm_k$, $D$

        \STATE $sort\_by\_market(BR)$ \label{alg:ws:sort1} \COMMENT{/*Prioritizes non-burstable on-demand VMs*/}
        
        \FOR{\textbf{each NON-burstable} $vm_j \in BR$}

                \STATE $S \leftarrow selectTasks(vm_j)$ \label{alg:work_steal:select} \COMMENT{/*Get tasks that can be stolen*/}
                
                \FOR{\textbf{each} $t_i \in S$} \label{alg:work_steal:alloc:start}

                    \IF{$check\_migration(t_i,vm_k, D)$} \label{alg:work_steal:alloc:check}
                        \IF{$vm_k$ \textbf{ is } burstable} \label{alg:work_stel:isburst}
                            \STATE $set\_baseline\_mode(t_i, vm_k)$\label{alg:migration_proc:set_base}
                            \STATE $migrate(t_i,vm_k)$ \label{alg:work_stel:burst}
                            \STATE \textit{stops the loop} \label{alg:work_steal:stop}
                
                        \ELSIF{$vm_k \textbf{ not }$ burstable}
                            \STATE $migrate(t_i,vm_k)$ \label{alg:work_stel:mig}
                        \ENDIF
                    \ENDIF
                    
                \ENDFOR \label{alg:work_steal:alloc:end}
                
        \ENDFOR
        
        \IF{\textbf{at least one task was stolen}}
            \STATE $BR \leftarrow BR \cup \{vm_k\}$ \label{alg:work_steal:alloc:busy}
            \STATE $IR \leftarrow IR \backslash \{vm_k\}$
            \label{alg:work_steal:remove:idle}
        \ENDIF

	\end{algorithmic}
\end{algorithm}
 }

\section{Experiments and Results}
\label{sec:experimental_tests}

According to AWS, only spot VMs of types C3, C4, C5, M4, M5, R3, and R4 with memory  amount smaller than 100 GB are hibernation-prone. Therefore, in our experiments, we used spot VMs C3 and C4 which are compute-optimized and have high availability in the spot market~\footnote{\url{https://aws.amazon.com/ec2/spot/instance-advisor/}}. Moreover, we also use the T3.large instances that are the newest generation of the general proposed burstable instances of EC2. Table \ref{VMSCharacteristics} shows the  computational  characteristics  and the corresponding prices  in on-demand and spot markets in September 2020. 

\begin{table}[!htbp]
\centering
\caption{VMs attributes \label{VMSCharacteristics}}
\resizebox{\columnwidth}{!}{\begin{tabular}{cccccc}
\hline
\multirow{2}{*}{\textbf{Type}} & \multirow{2}{*}{\textbf{\#VCPUs}} & \multirow{2}{*}{\textbf{Memory}} &\multicolumn{2}{c}{\textbf{Price per Hour}} & \textbf{Baseline}\\
              &                  &                 & on-demand & spot  & \textbf{performance}  \\
\hline
C3.large		& 2	 	& 3.75 GB 	& 0.105\$   & 0.0299\$ & -  \\

C4.large		& 2		& 3.75 GB 	& 0.100\$   & 0.0366\$  & - \\

C3.xlarge		& 4		& 7.50 GB 	& 0.199\$   & 0.0634\$  & - \\

T3.large		& 2		& 8 GB	    & 0.0832\$   & -    & 20\% \\

\hline
\end{tabular}}
\end{table}

In our experiments, we have considered two distinct workloads: i) synthetic BoT applications composed by tasks generated with the application template proposed in Alves \textit{et al.}~\cite{MELOALVES2017150}, and ii) the ED application available in the  GRIDNBP 3.1 suite of NAS benchmark \cite{bailey1995parallel}. The tasks of the synthetic applications execute vector operations whose execution times depend on the size of the vectors. Hence, several synthetic tasks are created, each one with a memory footprint between 2.81 MB and 13.19 MB, resulting in execution times which vary from 102 to 330 seconds. Then, we conceived three synthetic BoT applications, J60, J80, and  J100 by randomly selecting the tasks.  Concerning the ED application, a real embarrassingly distributed application, the job ED200 is composed of 200 tasks solving the largest problem size (class B).

Table \ref{tab:jobInfo} shows the characteristics of the four jobs, including their respective number of tasks and memory footprint. The same deadline $D=2700$ seconds (45 minutes) is considered for all evaluated jobs and experiments,  corresponding to the shortest deadline within which Burst-HADS can find a feasible solution for all evaluated jobs, respecting the constraints defined in Section \ref{sec:problem_definition}. 

\begin{table}[H]
    \centering
    \caption{Characteristics of the Jobs \label{tab:jobInfo}}
\resizebox{0.9\columnwidth}{!}{\begin{tabular}{ccccc}
    \toprule
    \multirow{2}{*}{\textbf{job}} &  \multirow{2}{*}{\textbf{\# tasks}} &  \multicolumn{3}{c}{\textbf{memory footprint}}\\
    & &   min & avg & max\\
    \midrule
    J60 & 	60  &  2.85MB & 4.69MB  &12.20MB\\
    J80 &	80  &  2.91MB & 4.71MB & 13.19MB\\
    J100 &	100 &  2.81MB & 4.49MB & 10.86MB\\
    ED200 & 200 &  153.74MB&    168.68MB   & 177.77MB \\
    \bottomrule
    \end{tabular}}
    
\end{table}

For the execution of the ILS Primary Task Scheduling (Algorithms \ref{alg:ils} and \ref{alg:localsearch}), the following input parameters were empirically determined:  $\alpha = 0.5$,  $max\_iteration = 200$; $max\_attempt = 50$; $swap\_rate = 0.10$; $max\_failed = 20$; $relaxed\_rate  = 0.25$; $burst\_rate = 0.2$. Moreover,  we defined  $AC=900$ seconds. As said in section \ref{sec:migration}, Burst-HADS uses a checkpoint approach proposed in \cite{teylocc}. In that approach, the user defines the parameter $ovh$, the maximum percentage of time overhead that the checkpoint mechanism is allowed to add in the original execution time of a task. From this parameter, Burst-HADS defines the number of checkpoints of each task and also the period between checkpoints. In this work, $ovh = 10\%$ for all tests.

We have firstly evaluated Burst-HADS in a scenario without hibernation, comparing it with (1) the schedule given by the proposed ILS  using only on-demand VMs and (2) the schedule given by  HADS  in a scenario without hibernation. The aim of these experiments is to measure the impact in the monetary cost and execution time of including  burstable on-demand VMs into the scheduling procedure. Table \ref{tab:baseline} presents the average of three executions of the synthetic jobs J60, J8, and J100,  and the real application ED200, for each case.

\begin{table}[htb]
\centering
\caption{Cost and Makespan of Burst-HADS and  HADS, without hibernation; and  ILS On-demand only.   \label{tab:baseline}}
\begin{tabular}{ccccccc}
\toprule
\multirow{2}{*}{\textbf{JOB}} & \multicolumn{2}{c}{\begin{tabular}[c]{@{}c@{}}\textbf{Burst-HADS} \\\textbf{ Without Hibernation}\end{tabular}} & \multicolumn{2}{c}{\begin{tabular}[c]{@{}c@{}}\textbf{HADS} \\
\textbf{Without Hibernation}\end{tabular}} & \multicolumn{2}{c}{\begin{tabular}[c]{@{}c@{}}\textbf{ILS On-demand} \end{tabular}} \\
\cline{2 - 7}
                     & cost    & makespan    & cost                                       & makespan                                   & cost                                  & makespan                              \\
\midrule
                
J60   & \$0.112      & 1274     & \$0.067  & 2290     & \$0.271  & 1112 \\
J80   & \$0.151      & 1329     & \$0.104  & 2295     & \$0.312  & 1190 \\
J100  & \$0.176      & 1660     & \$0.112  & 2332     & \$0.371  & 1462 \\
ED200 & \$0.357      & 2275	    & \$0.267 &	2580	  & \$0.698	 & 1887  \\
\bottomrule
\end{tabular}
\end{table}

  In comparison with HADS, Table \ref{tab:baseline} shows that  Burst-HADS reduces the makespan in 44.37\%,  42.09\%, 28.82\%, and 11.82\%, for jobs J60, J80, J100, and ED200, respectively. However, the average monetary cost increases by 66.34\%, 44.54\%, 57.55\%, and 33.71\%,  for the same comparison. The latter increases because Burst-HADS already starts by using some burstable on-demand VMs 
  Moreover, the ILS based primary scheduling uses more VMs to reduce the execution time,  while in HADS, the initial scheduling aims at minimizing only the monetary cost.


On the other hand, compared to the ILS on-demand strategy, on average,  Burst-HADS reduces the monetary cost by more than 52.00\%, with an average increase of 15\% in the makespan. The ILS on-demand strategy uses the scheduling plan given by the ILS proposed in Algorithm \ref{alg:ils}, which does not include spot neither burstable VMs, but only regular on-demand VMs.  The longer makespan is due to the execution of tasks in the baseline mode of burstable VMs, which does not occur in the ILS on-demand strategy. 

Since cloud users have no control over spot VM hibernations, we have emulated different patterns of spot VM hibernations and resuming times by using Poisson distribution \cite{ahrens1974computer} for each type of spot VM. We have applied distinct Poisson functions for modeling the events in order to have distinct scenarios where  \textit{hibernating} and \textit{resuming} events have different probability mass functions defined by the parameters $\lambda_h$ and $\lambda_r$, respectively. Let the $\lambda$ parameter of Poisson distribution be the number of expected events divided by a time interval.  Since the application execution is discretized by time interval and $D$ is the application deadline, if we respectively define $ k_h $ and $ k_r $,  as the expected number (rate) of  hibernating and resuming events during the application execution,   $\lambda_h$ and $\lambda_r$ parameters are given by $ \lambda_h = k_h/D $ and $\lambda_r = k_r/D$. We should point out that, since we are considering scenarios where multiple events can happen, the actual number of hibernation (resp., resuming) events that occur in an experiment might be greater than $ k_h$ (resp., $ k_r $). Table \ref{tab:poisson_lambda} presents five different scenarios by varying $k_h$ and $k_r$.

\begin{table}[ht]
    \centering
    \caption{Different execution scenarios generated by varying  parameters $\lambda_h$ and $\lambda_r$  \label{tab:poisson_lambda}}
\resizebox{0.9\columnwidth}{!}{    \begin{tabular}{ccccc}
    \toprule
    ID & \textit{hibernating} & \textit{resuming}  & $\lambda_h$ & $\lambda_r$ \\
    \midrule
    $sc_1$ & $k_h=1$ & $k_r=0$ &  $1/2700$ & $0/2700$\\
    $sc_2$ & $k_h=5$ & $k_r=0$ &  $5/2700$ & $0/2700$\\
    $sc_3$ & $k_h=1$ & $k_r=5$ &  $1/2700$ & $5/2700$\\
    $sc_4$ & $k_h=5$ & $k_r=5$   &  $5/2700$ & $5/2700$\\
    $sc_5$ & $k_h=3$ & $k_r=2.5$ &  $3/2700$ & $2.5/2700$\\ 
    \bottomrule
    \end{tabular}}
    
\end{table}

Table \ref{tab:result} presents the averages of three executions of jobs J60, J80, J100, and ED200 using Burst-HADS and HADS in each of the five execution scenarios.  For each job and scenario, the table shows the average number of hibernations followed by resume events. It also includes the number of non-burstable on-demand VMs launched by the Dynamic Module of both Burst-HADS and HADS that handle the hibernations as well as the average of monetary cost and makespan. Finally, the two last columns represent the percentage difference  between Burst-HADS and HADS (diff) related to the monetary cost and the makespan. 

\begin{table*}[htb]
\centering
\caption{Comparison between Burst-HADS and HADS in terms of monetary cost and makespan in scenarios $sc_1$ to $sc_5$ \label{tab:result}}
\resizebox{0.85\textwidth}{!}{\begin{tabular}{cccccccccccc}
\toprule
\multirow{2}{*}{ \textbf{Job}} &
\multirow{2}{*}{ \textbf{scenario} } &
\multirow{2}{*}{\textbf{ \# hibernations }} &
\multirow{2}{*}{ \textbf{\# resume }} &
\multicolumn{2}{c}{ \begin{tabular}[c]{@{}c@{}} \textbf{\# used regular on-demand VMs}\end{tabular}} &
\multicolumn{2}{c}{ \textbf{Burst-HADS}} & 
\multicolumn{2}{c}{\textbf{HADS}} & 
\multicolumn{2}{c}{\begin{tabular}[c]{@{}c@{}}  \textbf{Diff (\%)}  \end{tabular}} \\ 
\cline{5-12} 
&  &  &  & Burst-HADS  & HADS  &  cost &  makespan    &  cost  &  makespan & cost    &  makespan \\ \midrule
\multirow{5}{*}{J60}    & $sc_1$ &	0.66 &	0.00 &	0.00 &	0.00 &	\$0.119 &	1274 &	\$0.091 &	2620 &	-30.77\% &	   51.37\%\\
                        & $sc_2$ &	3.33 &	0.00 &	1.33 &	2.33 &	\$0.204 &	1277 &	\$0.257 &	2549 &	 20.54\% &	   49.90\%\\
                        & $sc_3$ &	2.33 &	2.33 &	1.33 &	0.00 &	\$0.127 &	1752 &	\$0.101 &	2539 &	-26.07\% &	   31.00\%\\
                        & $sc_4$ &	5.33 &	4.00 &	1.67 &	0.00 &	\$0.142 &	1857 &	\$0.119 &	2634 &	-19.90\% &	   29.50\%\\
                        & $sc_5$ &	2.66 &	1.00 &	1.33 &	2.00 &	\$0.150 &	1445 &	\$0.169 &	2359 &	 11.44\% &	   38.75\%\\
\midrule
\multirow{5}{*}{J80}    & $sc_1$ &	1.00 &	0.00 &  1.33	 & 0.33	  & \$0.167 &   1419 &   \$0.150 &	2581 &	-11.33\%  &	45.03\% \\
                        & $sc_2$ &	5.00 &	0.00 &	1.00     & 	3.00  &	\$0.210 & 	2267 &   \$0.298 &	2591 &	29.48\%   &	12.50\% \\
                        & $sc_3$ &	3.00 &	1.00 &	1.67     & 	1.00  &	\$0.164 & 	1367 &   \$0.147 &	2602 &	-11.34\%  &	47.46\% \\
                        & $sc_4$ &	9.66 &	7.66 &	1.00     & 	2.00  &	\$0.244 & 	2488 &   \$0.212 &	2607 &	-15.25\%  &	4.56\% \\
                        & $sc_5$ &	3.00 &	1.00 &	1.33     & 	3.00  &	\$0.195 &	1589 &   \$0.246 &	2529 &	20.47\%   &	37.17\% \\
\midrule
\multirow{5}{*}{J100}   & $sc_1$ &	2.00 &  0.00 &	0.00     &	0.00  &	\$0.191	&   1798 &	 \$0.157 &	2332 &	-21.76\%  & 22.90\% \\
    '                   & $sc_2$ &	7.00 &  0.00 &	1.33     &	3.00  &	\$0.212	&   1900 &	 \$0.353 &	2518 &	39.94\%	  & 24.54\% \\
                        & $sc_3$ &	6.00 &  3.00 &	1.67     &	1.00  &	\$0.201	&   1925 &	 \$0.166 &	2636 &	-21.08\%  & 26.97\% \\
                        & $sc_4$ &	11.00 &	9.00 &	1.00     &	0.00  &	\$0.286	&   2453 &	 \$0.278 &	2591 &	-2.88\%	  & 5.33\% \\
                        & $sc_5$ &	3.66  & 2.00 &	1.00     &	2.50  &	\$0.166	&   1547 &	 \$0.189 &	2543 &	12.49\%	  & 39.15\% \\
   
\midrule
\multirow{5}{*}{ED200} & $sc_1$ &	3.00 &	0.00 &	1.00 &	0.33 &	\$0.388 &	2327 &	\$0.314 &	2680 &	-23.57\% &	13.17\% \\
                        & $sc_2$ &	8.00 &	0.00 &	2.00 &	5.00 &	\$0.482 &	2448 &	\$0.512 &	2676 &	5.86\%   &	8.52\% \\
                        & $sc_3$ &	6.66 &	4.00 &	2.33 &	1.00 &	\$0.427 &	2345 &	\$0.387 &	2672 &	-10.34\% &	12.24\% \\
                        & $sc_4$ &	9.00 &	6.00 &	2.00 &	1.00 &	\$0.411 &	2560 &	\$0.389 &	2690 &	-5.66\%  &	4.83\% \\
                        & $sc_5$ &	4.33 &	2.33 &	1.67 &	3.00 &	\$0.367 &	2342 &	\$0.467 &	2674 &	21.41\%  &	12.42\% \\

\bottomrule
\end{tabular}}
\end{table*}

As can be observed in Table \ref{tab:result}, Burst-HADS minimizes the makespan in all execution cases, presenting an average reduction of 25.87\%.  As explained in Section \ref{sec:migration}, whenever a spot VM hibernates, Burst-HADS immediately migrates the interrupted tasks to other VMs. Thus,  the increment in the makespan is due to the overhead of this procedure, which can include the launch of new VMs. However, 
small jobs, i.e., jobs with fewer tasks,  are less affected on those scenarios than the biggest ones.  For example, while for job J60, the average makespan reduction, considering all scenarios, is 40.10\%,  for job ED200, that reduction is only 10.24\%. Such a behavior can be explained because, in our experiments, we have fixed the same deadline for all jobs and, therefore, small jobs have more spare time between its expected makespan defined by the ILS and the deadline. Consequently, in this case,  Burst-HADS benefits more from the burst mode of the burstable VMs since it has more idle time to earn CPU credits. Moreover, it also executes the work-stealing more frequently, which also reduces the makespan.

On the other hand,  independently of the scenario or job, HADS's makespan 
get closer to the deadline whenever a hibernation occurs. That happens because the HADS framework 
postpones as much as possible the execution of the migration procedure. Since HADS gives priority to the monetary cost save, its central idea is to wait for a hibernated  VMs resume and then avoid the launch of new VMs.

In scenarios $sc_2$ and $sc_5$, Burst-HADS improved the monetary cost for all  jobs. The $sc_2$ is the worst execution case scenario, since it has the highest rate of hibernation ($k_h=5$) and no resume rate ($k_r=0$), while $sc_5$ is the average case scenario where the rate of  hibernation is $k_h=3.0$, and the rate of resume is  $k_h=2.5$ (see Table \ref{tab:poisson_lambda}). In these scenarios, Burst-HADS uses fewer regular on-demand VMs than HADS. Moreover, in both cases,  the number of hibernations is higher than the number of resumes. Hence, compared to HADS, Burst-HADS is more efficient in minimizing the impact of spot hibernations in those cases by migrating tasks to busy and idle VMs and also to burstable VMs using the burst mode. It is worth pointing out that the average increase of Burst-HADS makespan was 1.92\%, considering all executions.

Compared to the ILS On-demand, both Burst-HADS and HADS minimized the monetary cost for all execution cases, presenting an average reduction of  41.80\% and 39.65\%, respectively. For job ED200, for example, the worst execution scenario, $sc_2$,  Burst-HADS reduced the monetary cost by 30.96\%, while HADS presented an average reduction of 26.66\%. 



\section{Conclusion and Future Work}
\label{sec:conclusion}

This paper proposes the Burst Hibernation-Aware Dynamic Scheduler (Burst-HADS), a dynamic scheduler that combines hibernation-prone spot VMs with burstable ones for executing bag-of-task applications subject to a deadline.  The objective of the scheduler is to minimize both the makespan and the monetary costs of the execution, respecting the application's deadline even when the spot VMs hibernate. Burst-HADS was evaluated in a real environment using VMs of AWS EC2, considering scenarios with different hibernation and resume rates. 

The results show that compared to the ILS On-demand approach that uses only regular on-demand VMs,  Burst-HADS can reduce the monetary cost for all execution scenarios at the expense of longer makespan due to the migration overhead and the baseline mode of the burstable VMs. Furthermore, compared to HADS, our previous proposed framework that uses only spot and on-demand VMs, Burst-HADS can reduce the makespan by more than 25 \%, with an average increment of only 1.92\% in the monetary cost. As future work, we intend to include in the migration decision of Burst-HADS some predictions about the future state of the spot market. For example, to consider which instance type would improve the reliability of the execution. A second research direction will be to conduct a study in order to determine which type of burstable VM  would offer the best tradeoff between price and performance in Burst-HADS. 

\section*{Acknowledgments}
This research was supported by \textit{Programa Institucional de Internacionaliza\c{c}\~{a}o} (PrInt) from CAPES as part of the project REMATCH (process number 88887.310261/2018-00).

\bibliographystyle{IEEEtran}
\bibliography{ipdps}

\end{document}